\documentclass[%
 showpacs,
 reprint,
 groupedaddress,
 showpacs,
 amsmath,amssymb,
 aps,
prb,
]{revtex4-1}

\usepackage[pdftex]{graphicx, color}
\usepackage{multirow,booktabs}
\usepackage{bm}
\usepackage{braket}
\usepackage{siunitx}
\let\Im\relax

\DeclareMathOperator{\Im}{Im}
\DeclareMathOperator{\Tr}{Tr}
\DeclareMathOperator{\tr}{tr}

\newcommand{\Table}[1]{Table~\ref{#1}}
\newcommand{\Eq}[1]{Eq.~\eqref{#1}}

\newcommand{\Fig}[1]{Fig.~\ref{#1}}
\newcommand{\Ref}[1]{Ref.~\cite{#1}}
\newcommand{\Sec}[1]{Sec.~\ref{#1}}
\newcommand{\Cite}[1]{~\cite{#1}}
\newcommand{\App}[1]{Appendix.~\ref{#1}}

\begin{document}

\title{Chirality polarizations and spectral bulk-boundary correspondence}

\author{Akito Daido}
\email[]{daido@scphys.kyoto-u.ac.jp}
\affiliation{%
 Department of Physics, Graduate School of Science, Kyoto University, Kyoto 606-8502, Japan
}%

\author{Youichi Yanase}
\affiliation{%
 Department of Physics, Graduate School of Science, Kyoto University, Kyoto 606-8502, Japan
}%


\date{\today}

\begin{abstract}
Surface physics dominated by bulk properties has been one of the central interests in modern condensed matter physics, from electric polarization to bulk-boundary correspondence of topological insulators and superconductors.
Here, we extend theory of electric polarization to \textit{chirality polarizations}, that is, surface charges corresponding to local antisymmetries characterized as a bulk property.
Using the notion of chirality polarizations, we prove the recently proposed spectral bulk-boundary correspondence, a generalization of bulk-boundary correspondence in chiral symmetric systems into complex frequencies.
We show a {physically transparent proof} via Wannier functions and a formal proof by considering the adiabatic change of surface chirality charges, highlighting the similarities and the differences between electric polarization and chirality polarizations.
\end{abstract}



\maketitle


\section{Introduction}
Bulk-boundary correspondence (BBC) is one of the central topics in modern condensed matter physics.
It has been clarified that existence of surface states is ensured by nontrivial topological invariants characterized solely by bulk properties\Cite{Chiu2016}.
BBC predicts non-dissipative surface spin currents and Majorana zero modes in topological insulators and superconductors, respectively. {They might be promising building blocks for spintronics} and topological quantum computation\Cite{Qi2011,Pesin2012,Sato2016review}.
Thus, BBC is an important concept both from fundamental and practical points of view.

There is another context where surface physics is dominated by the bulk.
Electric polarization, or the accumulated surface charge, is predicted by the Berry phase formula up to a polarization quantum
\Cite{KingSmith1993,Vanderbilt1993,Resta1994,Vanderbilt2018}.
Naively speaking, electric polarization is given by the expectation value of the position operator in open boundary conditions (OBC): $p=\braket{\hat{x}}_{\mathrm{OBC}}/L$ {}{(in one dimension)}. This expression apparently picks up contribution from the surface by $O(1)$.
It is quite amazing in this sense that electric polarization is a bulk geometric property.
Similarly, the orbital magnetic dipole moment{}{, which is  expressed as }
${m}_{\mathrm{orb}}=\braket{\hat{{x}}\hat{j}_y-\hat{y}\hat{j}_x}_{\mathrm{OBC}}/2L^2$ {}{in two dimensions,}
 is also known to be characterized by the bulk, and {a formula evaluating ${m}_{\mathrm{orb}}$ in periodic boundary conditions (PBC) has been obtained\Cite{Vanderbilt2018,Thonhauser2005,Ceresoli2006,Souza2008}.}
These formulas imply a kind of BBC, in the sense that quantities apparently sensitive to surface information are determined by the bulk.
{In the following, we refer to such relations as geometric BBC, in analogy with topological BBC.}

In contrast to electric and magnetic dipole moments, physical nature of higher order multipole moments is hardly understood, and PBC formulas have been lacking.
{For example, electric quadrupole and octupole moments
can be determined by the bulk when certain crystalline symmetries exist\Cite{Benalcazar2017a,Benalcazar2017b}, but extension to general situations seems to be still an ongoing issue\Cite{Kang2018,Wheeler2018,Ono2019}}.
There is also an attempt to characterize spin magnetic quadrupole moments {such as $M_{xy}=\braket{\hat{x}\hat{s}_y+\hat{y}\hat{s}_x}_{\mathrm{OBC}}/{L^2}$} {}{(for two-dimensional systems)}, but resultant expressions are gauge-dependent\Cite{Chen2018}, implying they might not be a bulk property.
Magnetic quadrupole moments have also been discussed from the viewpoint of local thermodynamics, and the obtained gauge-invariant formulas are directly related to the magnetoelectric effect and intriguing transport phenomena\Cite{Gao2018a,Gao2018b,Shitade2018a,Shitade2019}.
{However, their relationship with the quantities such as $M_{xy}$ has not been clarified.}
Multipole moments are fundamentally important quantities because they contribute to electromagnetic fields as coefficients of the multipole expansion\Cite{Jackson}.
Therefore, it is an important issue to gain deeper insight into for what physical quantities ``BBC" holds, along with their relation to local thermodynamics.

\begin{figure}
\centering
\includegraphics[width=8cm]{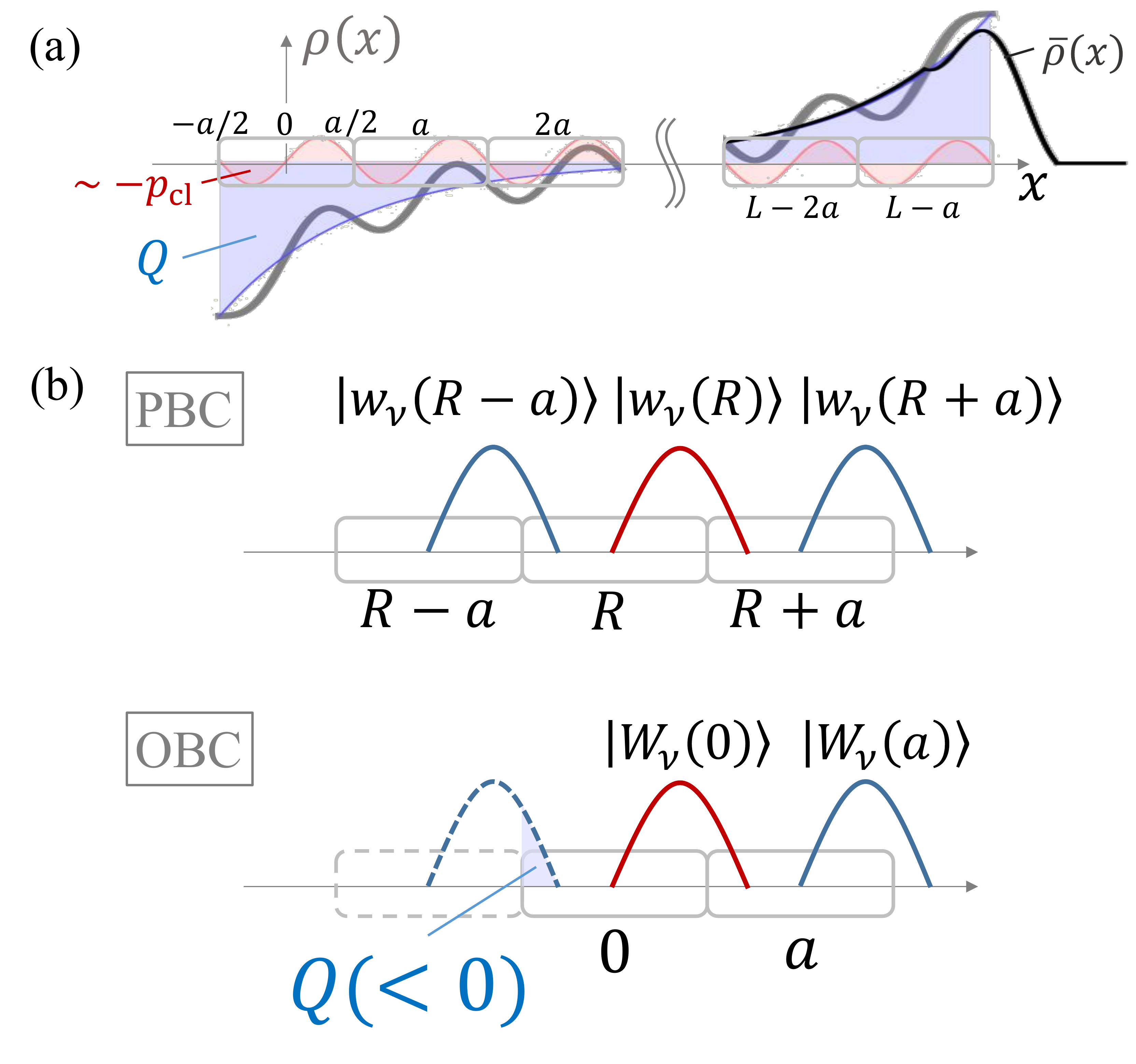}
\caption{(a) A schematic figure of electric charge density of the system under considerations. We show $\bar{\rho}(x)$ only around $x=L$.
Similar figures are obtained for chirality charges by replacing $\rho(x)$ with $\rho_f(x,z)$.
(b) Intuitive picture for the origin of microscopic excess charge bound at the left end\Cite{Rhim2017}. In Figs.\ref{fig:fig1}(a) and \ref{fig:fig1}(b), we ignored surface modification of GWFs from the Wannier functions in PBC, for simplicity.
}
\label{fig:fig1}
\end{figure}

Recently, Tamura et al. proposed an interesting relation in chiral symmetric one-dimensional systems\Cite{Tamura2018}:
surface accumulation of a component of Green's function proportional to the chiral operator $\Gamma$ coincides with a generalization of the winding number in the bulk; That is, the equality $F(z)=w(z)/z$ holds, with
\begin{subequations}
\begin{gather}
F(z)=\sum_s\int_{-a/2}^{R_c-a/2}dx\,\braket{x,s|\Gamma G(z)|x,s},\label{eq:F}\\
w(z)\equiv-\int\frac{dk}{4\pi i}\tr\left[\Gamma G_k(z)\partial_k G_k^{-1}(z)\right],\label{eq:w}
\end{gather}
\end{subequations}
where $z\in\mathbb{C}$ is off the energy bands on the real axis.
In \Eq{eq:F}, $G(z)=(z-H)^{-1}$ is the Green's function and $H$ is the Hamiltonian in OBC.
The integral runs from the left end of the system $x=-a/2$ to a certain bulk unit cell whose left end is $R_c-a/2$ [see Fig.~\ref{fig:fig1}]. As we see later, \Eq{eq:F} is independent of the choice of $R_c$ as long as it is deep inside the bulk.
In \Eq{eq:w}, $G_k(z)=(z-H_k)^{-1}$ and $H_k=e^{ik\hat{R}}He^{-ik\hat{R}}$, with $H$ the Hamiltonian in PBC.
The definition of $\ket{x,s}$ and $\hat{R}$ is shown later.
{}{Thus, the equivalence of \Eq{eq:F} and \Eq{eq:w} reveals a BBC.}

The authors named the relation \textit{spectral bulk-boundary correspondence} (SBBC) after the usual BBC reproduced in the limit $z\to 0$ (note that $w(0)$ is the {topological} winding number\Cite{Essin2011,Sato2011,Gurarie2011}).
{}{Investigation of this novel BBC would not only give us a deeper understanding of BBC, but also tell us how to deal with position operators in periodic crystals (as we see later).}
In the context of superconductivity, SBBC identifies as a bulk property surface accumulation of odd-frequency Cooper pairs (since $F(-z)=-F(z)$), which plays an essential role in the superconducting hetero-structures and Josephson junctions\Cite{Bergeret2005,Tanaka2012}.
Reference~\cite{Tamura2018} addressed numerical evidences supporting SBBC for various systems including the Kitaev chain\Cite{Kitaev2001} and the Majorana nanowire\Cite{Lutchyn2010,Oreg2010}, which are the central platform for topological superconductivity\Cite{Sato2016review}.
For some special parameters of Kitaev chain, SBBC was also analytically illustrated\Cite{Tamura2018}.
However, formal proof of SBBC has not been given.

In this paper, we present a formal proof of SBBC.
{}{We consider only one-dimensional systems, while our results are straightforwardly applicable to higher-dimensional systems with clean surfaces by fixing a k-point in the surface Brillouin zone.}
Our derivation is based on an analogy with electric polarization;
That is, SBBC can be regarded as a condensed notation of topological and geometric BBC for a series of \textit{chirality polarizations}.
We thus add chirality polarizations to members of physical quantities where BBC holds.

The following part of the paper is constructed as follows.
{}{In Sec.~\ref{sec:2}}, we briefly review the derivation of electric polarization in OBC\Cite{KingSmith1993,Vanderbilt1993,Resta1994,Vanderbilt2018,Rhim2017}.
{}{In Sec.~\ref{sec:chiralitypol}}, we discuss how analogy between electric and chirality polarizations is established, and address a physical explanation for SBBC via Wannier functions with some natural assumptions.
{}{In Sec.~\ref{sec:comment}}, we comment on the difference from the spin magnetic quadrupole moment, or the spin density polarization, for which a gauge-dependent result has been reported with Wannier-function based formalisms\Cite{Thole2016,Chen2018}.
{}{Finally in Sec.~\ref{sec:derivation}}, we show a formal proof of SBBC based on the adiabatic deformation, in analogy with the fact that change of electric polarization is given by the transient current through the bulk region.
No assumption is required for the latter proof, and thus we complete the derivation of SBBC.
{}{This section can be read separately from other sections.
We provide a brief conclusion in Sec.~\ref{sec:conclusion}}

\section{A review of electric polarization}
\label{sec:2}
In this section, we review derivation of electric polarization and its relation to surface charge.
The notations introduced below are summarized in \App{sec:notations}.
\subsection{electric polarization and Berry phase}
We first review the derivation of Berry phase formula for electric polarization in OBC\Cite{KingSmith1993,Vanderbilt1993,Resta1994,Vanderbilt2018}.
Let us consider a one-dimensional sample spreading over $-a/2\le x< L-a/2$, with $a$ the lattice constant (\Fig{fig:fig1}(a)).
The shift $-a/2$ is just a technical simplification, and is not essential.
{}{The system is assumed to be gapped, and following Refs.~\cite{Vanderbilt1993,Vanderbilt2018}, it is assumed that the space of occupied electron states can be spanned by generalized Wannier functions (GWFs) $\ket{W_\nu(R)}$ satisfying the following properties: (1) they are exponentially localized around each lattice point $R$, and (2) they are asymptotically equivalent with Wannier functions in PBC, as $R$ getting away from the surface\Cite{Rehr1974}.
{Existence of such GWFs has been proved at least in the absence of spin-orbit coupling\Cite{Rehr1974,Kivelson1982,Nenciu1998}.}
Here, the index $\nu$ specifies GWFs associated with the same unit cell, and thus $\sum_\nu1$ is the number of electrons within a unit cell.
We also assume, for simplicity, ions to be classical point charges $Z_i$ located at $R+u_i(R)$, and set the electron charge to be unity.}

Electric polarization is defined by
\begin{equation}
p\equiv\frac{1}{L}\int_{-a/2}^{L-a/2} dx\,x\rho(x),\quad \rho(x)=\rho_{\mathrm{el}}(x)+\rho_{\mathrm{ion}}(x).\label{eq:pol2}
\end{equation}
Here, $\rho(x)$ is the microscopic electric charge density which vanishes outside of the sample.
Contributions from electrons and ions are represented by $\rho_{\mathrm{el}}(x)$ and $\rho_{\mathrm{ion}}(x)$, respectively:
\begin{subequations}
\begin{gather}
\rho_{\mathrm{el}}(x)=\sum_{R,s,\nu}|\braket{x,s|W_\nu(R)}|^2,\label{eq:rho_el}\\
\rho_{\mathrm{ion}}(x)=\sum_{R,i}Z_i\delta(x-R-u_i(R)),\label{eq:rho_ion}
\end{gather}
\end{subequations}
where the index $s$ runs over the internal degrees of freedom such as spin, including the Nambu degree of freedom when superconductors are concerned in \Sec{sec:chiralitypol}.
The ket $\ket{x,s}$ is the position eigenstate with $x$ taking continuous values in $[-a/2,L-a/2)$,
and the position operator is written as
\begin{equation}
\hat{x}=\sum_s\int_{-a/2}^{L-a/2}dx\,x\,\ket{x,s}\bra{x,s}.
\end{equation}
As is well known, origin independence of the polarization is ensured by the neutrality of the total charge,
\begin{equation}
  \int_{-a/2}^{L-a/2}dx\,\rho(x)=N\left(\sum_\nu\,1+\sum_iZ_i\right)=0,\label{eq:tcn}
\end{equation}
with $N=L/a$ the total number of unit cells.

By substituting Eqs.~\eqref{eq:rho_el} and \eqref{eq:rho_ion} into Eq.~\eqref{eq:pol2}, we obtain
\begin{subequations}
\begin{align}
p&=\frac{1}{L}\sum_{R,\nu}\braket{W_\nu(R)|\hat{x}|W_\nu(R)}+\frac{1}{L}\sum_{R,i}Z_i(R+u_i(R))\\
&=\frac{1}{L}\sum_{R,\nu}\braket{W_\nu(R)|(\hat{x}-R)|W_\nu(R)}+\frac{1}{L}\sum_{R,i}Z_iu_i(R).\label{eq:pol3}
\end{align}
\end{subequations}
Here we have used the charge neutrality condition \Eq{eq:tcn} to obtain the second line.

Equation~\eqref{eq:pol3} is essential to pass to PBC. The dominant contribution to the first term comes from the lattice points in the bulk, since each term in the summand is $O(1)$ due to localization of GWFs.
Thus, we have only to take bulk contribution to obtain $p$ in the thermodynamic limit.
By assumption, GWFs in the bulk can be replaced with the genuine Wannier functions $\ket{w_\nu(R)}$ in PBC.
Thus, electronic contribution to the polarization is given by the Berry phase formula\Cite{KingSmith1993,Vanderbilt1993,Resta1994,Vanderbilt2018},
\begin{subequations}
\begin{align}
p_{\mathrm{el}}&=\frac{1}{a}\sum_\nu\braket{w_\nu(0)|\hat{x}|w_\nu(0)}\label{eq:BerryPhaseFormula}\\
&=i\sum_\nu\int\frac{dk}{2\pi}\braket{u_\nu(k)|\partial_k u_\nu(k)},\label{eq:BerryPhaseFormula2}
\end{align}
\end{subequations}
while ionic contribution is given by $p_{\mathrm{ion}}=\sum_i Z_iu_i(0)$.
Here, $\ket{w_\nu(R)}$ and $\ket{u_\nu(k)}$ are defined through the Bloch waves whose gauge are taken to be periodic in $k$, $\ket{\psi_\nu(k)}=\ket{\psi_\nu(k+2\pi)}$, by
\begin{subequations}
\begin{gather}
\ket{u_\nu(k)}=e^{-ik\hat{x}}\ket{\psi_\nu(k)},\\
\ket{w_\nu(R)}=\frac{1}{\sqrt{N}}\sum_{k}e^{-ikR}\ket{\psi_\nu(k)},\label{eq:Wannier_def}
\end{gather}
\end{subequations}
when the bands are isolated with each other.
It should be noted that $\ket{u_\nu(k+2\pi)}=e^{-2\pi i\hat{x}}\ket{u_\nu(k)}$ is not equivalent to $\ket{u_\nu(k)}$\Cite{Vanderbilt2018} in this gauge choice.
For multiband cases, $\ket{\psi_\nu(k)}$ should be understood as a unitary-transformed energy eigenstate within the space of occupied states\Cite{Marzari1997,Marzari2012} [see also \App{sec:SBBCcalc}].

\subsection{Electric polarization and surface charge}
\label{sec:sec2B}
Next, we summarize the relation between the electric polarization and the surface charge.
{}{As discussed in classical electromagnetism, electric polarization should be equivalent to the surface charge density. This subsection is devoted to establish a quantitative description of this relation in quantum mechanics.}

{}{In Fig.~\ref{fig:fig1}(a), we illustrated a typical charge distribution in a system.
First, as highlighted by red, charge density generally oscillates within a unit cell even in the bulk.
When its distribution has a polar character as shown in Fig.~\ref{fig:fig1}(a), it gives a contribution to electric polarization and surface charge.
This contribution would be regarded as classical contribution, and therefore we write this as $p_{\mathrm{cl}}$.}

{}{It should also be noted that $p_{\mathrm{cl}}$ is not the only contribution to surface charge.
In addition to this charge asymmetry within a unit cell, there might exist a genuine extra charge\Cite{Rhim2017} around the surface as highlighted by blue.
The origin of this extra charge $Q$ can be understood from Fig.\ref{fig:fig1}~(b).
Let us consider the simplest case where only two Wannier functions (in PBC) contribute to the charge density within a unit cell (the upper panel).
When we terminate the system, the unit cell at the left end loses charge neutrality due to the lack of the left-neighboring unit cell and its Wannier function (the lower panel).
This gives a net excess charge $Q$.
Essentially the same discussion can be made even when modification of GWFs around the surface is taken into account\Cite{Vanderbilt1993,Rhim2017}.
%
%
}

{}{Quantitative description of this idea has been established well.
The equivalence of the electric polarization $p$ calculated in the previous section with $p_{\mathrm{cl}}-Q$ is shown with two steps.
First, $p$ is equivalent to $-\sigma_B$, with $\sigma_B$ the coarse-grained surface charge bound at the left edge\Cite{Vanderbilt1993,Vanderbilt2018,Baldereschi1988,Kudin2007,Rhim2017}:
\begin{equation}
\sigma_B\equiv\int^{x_c}_{-3a/2}dx\, \bar{\rho}(x),\quad \bar{\rho}(x)\equiv\frac{1}{a}\int_{-a/2}^{a/2}dr\,\rho(x+r),\label{eq:pol1}
\end{equation}
where $\bar{\rho}(x)$ stands for coarse-grained charge density.
Importantly, Eq.~\eqref{eq:pol1} is independent of the cutoff $x_c$ as long as it is deep inside the sample,
since
\begin{equation}
  \bar{\rho}(x)=0,\quad (x\in \mathrm{bulk})\label{eq:lcn}
\end{equation}
due to charge neutrality within a unit cell.}
In the following, the upper and lower bounds of the integral are sometimes denoted as $\pm\infty$ for simplicity, since $\rho(x)$ and $\bar{\rho}(x)$ vanish identically for $x$ far away from the sample.
With this notation, the relation $p=-\sigma_B$ is shown as follows:
\begin{subequations}
  \label{eq:polcharge}
\begin{align}
\frac{1}{L}\int^{\infty}_{-\infty}dx\,x\rho(x)&=\frac{1}{L}\int^{\infty}_{-\infty}dx\,x\bar{\rho}(x)\\
&=\int_{x\sim L}dx\,\bar{\rho}(x)+O(1/L)\\
&=-\int_{x\sim 0}dx\,\bar{\rho}(x)+O(1/L)\\
&=-\int_{-\infty}^{x_c}dx\,\bar{\rho}(x)+O(1/L).
\end{align}
\end{subequations}
We used the charge neutrality\Eq{eq:tcn} to derive the first and the third equalities.
The second and final equalities follow from \Eq{eq:lcn}.
We do not write $O(1/L)$ corrections explicitly in the following.

{}{Second, $\sigma_B$ is equivalent to $-p_{\mathrm{cl}}+Q$\Cite{Rhim2017}.
In order to get an intuition of this relation, we show a possible position dependence of $\bar{\rho}(x)$ in \Fig{fig:fig1}(a) around $x\sim L$ (shown with black line).
In contrast to the bare charge density $\rho(x)$ (shown with gray line), $\bar{\rho}(x)$ takes finite values for $L-a/2<x\le L$,
in particular picking up the classical polarization charge $p_{\mathrm{cl}}$.
We see a similar distribution of $\bar{\rho}(x)$ for $-a<x\le -a/2$.
Furthermore, $\bar{\rho}(x)$ naturally includes contribution from $Q$.
This is the reason why $\sigma_B=-p_{\mathrm{cl}}+Q$ holds.}

{}{Microscopically, classical contribution $p_{\mathrm{cl}}$ can be defined as follows:
\begin{equation}
p_{\mathrm{cl}}=\frac{1}{a}\int_{-a/2}^{a/2}dr\,r\rho(R_c+r),
\end{equation}
with $R_c=N_ca$ ($N_c\gg 1$) a lattice point in the bulk.
Ionic contribution is included to this term, since they are assumed to be point charges.
On the other hand, the genuine extra charge $Q$ is defined as
\begin{equation}
Q=\int_{-a/2}^{R_c-a/2}\!\!\!\!dx\,\rho(x).
\label{eq:Q}
\end{equation}
Compared with $p_{\mathrm{cl}}$, the contribution by $Q$ is quantum-mechanical, since $Q$ comes from the fact that electrons in crystals, or Wannier functions, are not point charges but are spreading over several lattice constants.
For derivation of the equality $\sigma_B=-p_{\mathrm{cl}}+Q$, we refer to Appendix B of Ref.~\cite{Rhim2017}.}

Here we note that the equalities $p=-\sigma_B$ and $\sigma_B=-p_{\mathrm{cl}}+Q$ are proven by using \textit{only the charge neutrality conditions} Eqs.~\eqref{eq:tcn} and \eqref{eq:lcn}\Cite{Rhim2017}.
This point is important for the discussion in \Sec{sec:chiralitypol}, where neutrality of chirality charges holds and plays an essential role in an analogous way.

To be precise, the discussion in this section ensures that surface charge is evaluated with the Berry phase only up to $O(1/L)$, in return for the simpleness of derivation. It is known that the correspondence is more precise, and the Berry phase correctly gives surface charge up to exponentially small finite-size correction\Cite{Vanderbilt1993,Vanderbilt2018}.
Similar situation also holds for chirality polarizations, as discussed later.

\subsection{{}{Genuine extra charge $Q$}}
It has been pointed out that the genuine extra charge $Q$ coincides with the inter-cellular Zak phase
\begin{equation}
Q=-i\sum_\nu\int\frac{dk}{2\pi}\braket{{U}_\nu(k)|\partial_k|{U}_\nu(k)},
\end{equation}
and captures dependence of topological surface states on surface terminations\Cite{Rhim2017}.
Here, $\ket{{U}_\nu(k)}$ is the periodic function $\ket{U_\nu(k+2\pi)}=\ket{U_\nu(k)}$ defined by
\begin{subequations}
\begin{gather}
\ket{{U}_\nu(k)}\equiv e^{-ik\hat{R}}\ket{\psi_\nu(k)},\\
\hat{R}\equiv\sum_{R,s}R\int^{a/2}_{-a/2} dr\,\ket{R+r,s}\bra{R+r,s}.
\end{gather}
\end{subequations}
The two definitions for periodic part of Bloch waves are connected by
\begin{subequations}
\begin{gather}
\ket{u_\nu(k)}=e^{-ik\Delta\hat{r}}\ket{U_\nu(k)},\\
\Delta\hat{r}\equiv\sum_{R,s}\int^{a/2}_{-a/2} dr\,r\ket{R+r,s}\bra{R+r,s}.
\end{gather}
\end{subequations}
The definition $\ket{U_\nu(k)}$ is sometimes used for calculation of topological invariants, since corresponding Bloch Hamiltonian $H_k=e^{-ik\hat{R}}He^{ik\hat{R}}$ satisfies the periodicity of the Brillouin zone: $H_{k+2\pi}=H_k$: For example, $H_k$ is used in the definition of the winding number, as well as its generalization \Eq{eq:w}.
Note that unit cell boundary is necessarily identified by introducing the unit-cell position operator $\hat{R}$.
This makes clear contrast to $\hat{x}$, which describes the absolute position independent of unit-cell choices.
Here, we choose $\hat{R}$ so as to reproduce the surface termination specified by OBC in question, and thus $H_k$ naturally includes the information of the surface termination.
This is why termination-dependent surface states can be predicted by $H_k$-based topological invariants such as the intercellular Zak phase in the presence of relevant symmetries.
An illustrative example of a system with three atoms in a unit cell is discussed in Ref.~\cite{Rhim2017}.

Finally, we point out that $Q$ can be concisely rewritten as
\begin{equation}
Q=-\frac{1}{a}\sum_\nu\braket{w_\nu(0)|\hat{R}|w_\nu(0)},\label{eq:eqQ}
\end{equation}
for the latter use
\footnote{To be precise, $\ket{w_\nu(0)}$ should be understood as the Wannier function at a home unit cell (such as $R=R_c$) far from the seam in PBC specified by $R=0$ and $L$, while $\hat{R}$ and $\hat{x}$ are the coordinate operators measured from there.}.
This expression directly expresses the illustration of $Q$ shown in Fig.~\ref{fig:fig1}(b).
\section{Proof of SBBC via GWFs}
\label{sec:chiralitypol}
In this section, we derive SBBC via GWFs and provide it a physical explanation.
We first introduce polarizations as well as charges of chirality, and then clarify the analogy between electric and chirality polarizations.
The following discussion holds as long as chiral symmetry exists.
Thus, examples include both superconducting systems and insulators with sublattice symmetries.
\subsection{Chirality polarization and chirality charge}
In order to get an intuition, let us expand \Eq{eq:w} around $z=0$ and concentrate on the coefficients:
\begin{subequations}
\begin{gather}
w(z)/z=\sum_{n=0}^\infty w_n\, z^{2n-1},\\
w_n\equiv-\int\frac{dk}{4\pi i}\tr\left[\Gamma H_k^{-(2n+1)}\partial_kH_k\right].\label{eq:an}
\end{gather}
\end{subequations}
A naive substitution of $\partial_kH_k=-i[\hat{R},H_k]$ into \Eq{eq:an} yields
\begin{equation}
w_n\sim\tr\bigl[\hat{e}_n\,\hat{R}\,\theta(-H)\bigr],\quad \hat{e}_n\equiv 2\Gamma H^{-2n},
\end{equation}
with $\theta(x)$ the Heaviside step function.
Similarity of $w_n$ to electric polarization is now clear: $w_n$ is almost equivalent to electric polarization when $\hat{e}_n$ is replaced with unity (electric charge). It would be appropriate to call $w_n$ the polarization of the $n$-th chirality charge $\hat{e}_n$, or the \textit{$n$-th chirality polarization}, since $\hat{e}_n$ is a quasi-local operator anticommuting with $H$. Note, however, that $\hat{R}$ appears instead of $\hat{x}$, in contrast to electric polarization $p_{\mathrm{el}}$ (contributed by electrons). In this sense, $w_n$ is more analogous to the accumulated microscopic charge $Q$ than to the physical polarization $p_{\mathrm{el}}$\footnote{{We can also establish another SBBC corresponding to $p_{\mathrm{el}}$, although we do not discuss it in detail.}}.
{}{We also refer the readers to Ref.\Cite{Shiozaki2013}, where similar discussion is made for $n=0$, as ``chiral polarization".}

With these observations in mind, let us define surface accumulation of the chirality charge density in OBC.
We rewrite \Eq{eq:F} as
\begin{equation}
F(z)\equiv\int_{-a/2}^{R_c-a/2}\!\!\!\!dx\,\rho_f(x,z),\quad (N_c=R_c/a\gg 1),\label{eq:def_Fz}
\end{equation}
with use of local chirality charges $\rho_f(x,z)$ defined by
\begin{equation}
\rho_f(x,z)\equiv\sum_s\braket{x,s|\Gamma G(z)|x,s},\quad G(z)\equiv(z-H)^{-1}.
\end{equation}
Note that the integration interval is the same as that of $Q$ (Eq.~\eqref{eq:Q}).
It should also be noted that $F(z)$ is the component of Green's function proportional to $\Gamma$ accumulated around the left end.

\subsection{Chirality-charge nuetrality conditions}
The ``charge neutrality conditions'' are also satisfied for chirality charges.
Let us first show the local neutrality condition corresponding to \Eq{eq:lcn}.
We consider coarse-grained local chirality charge density:
\begin{subequations}
\label{eq:cn_bulk}
\begin{align}
&\bar{\rho}_f(x,z)\equiv\frac{1}{a}\int^{a/2}_{-a/2}dr\,\rho_f(x+r,z)\\
&=\sum_{s}\frac{1}{a}\int^{x+a/2}_{x-a/2}dr\,\braket{r,s|\Gamma G(z)|r,s}.
\end{align}
\end{subequations}
When $x$ is sufficiently away from the surface, the integrand can be replaced with the corresponding quantity in PBC.
Thereby, we can take advantage of the periodicity of the lattice, to obtain
\begin{widetext}
\begin{equation}
\bar{\rho}_f(x,z)
\overset{\mathrm{OBC}\to\mathrm{PBC}}{\simeq}\frac{1}{N}\sum_{R,s}\frac{1}{a}\int^{x+a/2}_{x-a/2}dr\,\braket{R+r,s|\Gamma G(z)|R+r,s}.
\end{equation}
\end{widetext}
The righthand side is proportional to the trace of $\Gamma G(z)$ in PBC.
Thus, by expanding it with energy eigenstates, we obtain for $x$ in the bulk,
\begin{equation}
\bar{\rho}_f(x,z)\simeq\frac{1}{N}\sum_{k,\nu}\frac{1}{z-E_\nu(k)}\braket{\psi_\nu(k)|\Gamma|\psi_\nu(k)}=0.
\label{eq:lcn_f}
\end{equation}
This is the neutrality condition analogous to \Eq{eq:lcn}.

In the same way, total charge neutrality is also satisfied,
\begin{equation}
\int_{-\infty}^\infty dx\,\rho_f(x,z)=\sum_{E,n}\frac{\braket{E,n|\Gamma|E,n}}{z-E}=0,\label{eq:cn_tot}
\end{equation}
where $\ket{E,n}$ specifies energy eigenstates in OBC.
The label $n$ distinguishes, if any, degenerate eigenstates.
These charge neutrality conditions establish clear analogy with electric polarization.
This point is essential for the following proof.

{}{Before closing this section, we point out that an equation of continuity holds for local chirality charge density.
Let us assume that Hamiltonian is dependent on some parameter $\lambda$.
Then, we obtain
\begin{equation}
\int_{-\infty}^\infty dx\,\frac{\partial}{\partial\lambda}\rho_f(x,z)=0.
\end{equation}
This means that the presence of a chirality current $j_\Gamma(x)$ satisfying the equation of continuity
\begin{equation}
  \frac{\partial}{\partial\lambda}\rho_f(x,z)+\partial_xj_\Gamma(x)=0.
\end{equation}
The lattice version of this idea is used in Sec.~\ref{sec:derivation}.
}

\subsection{Proof of SBBC}
Here we provide a proof of SBBC by comparing chirality charges in OBC with chirality polarizations in PBC for each order $n$.
Let us carry out Laurent expansion of $\rho_f(x,z)$:
\begin{equation}
\rho_f(x,z)=\sum_{n=0}^\infty \rho_n(x)\,z^{2n-1}.
\end{equation}
After some algebra [see \App{sec:SBBCcalc}], we obtain
\begin{subequations}
\begin{gather}
\rho_0(x)=\sum_s\braket{x,s|\Gamma\hat{P}_{0}|x,s},\label{eq:index}\\
\rho_n(x)=\sum_s2\braket{x,s|\Gamma H^{-2n}\hat{P}_{\mathrm{occ}}|x,s}\ \ (n\ge1).\label{eq:rho_n}
\end{gather}
\end{subequations}
In the first line, $\hat{P}_0$ represents the projection operator onto (if any) gapless end states:
\begin{equation}
\hat{P}_0=\sum_{n}\ket{E=0,n}\bra{E=0,n}.
\end{equation}
In \Eq{eq:rho_n}, $\hat{P}_{\mathrm{occ}}$ stands for the projection operator onto occupied states where gapless end states are not included:
\begin{equation}
\hat{P}_{\mathrm{occ}}=\sum_{E<0,n}\ket{E,n}\bra{E,n}.
\end{equation}
The expansion of $\rho_f(x,z)$ naturally leads to the expansion of $F(z)$,
\begin{subequations}
\begin{gather}
F(z)=\sum_{n=0}^\infty Q_nz^{2n-1},\\
Q_n=\int_{-a/2}^{R_c-a/2}dx\,\rho_n(x).\label{eq:Q_n}
\end{gather}
\end{subequations}

Let us show the equivalence of chirality charges $Q_n$ with chirality polarizations $w_n$, from which SBBC immediately follows.
First, the $0$-th chirality charge reproduces topological BBC.
Actually, spatial integration of $\rho_0(x)$ gives the accumulated $0$-th chirality charge,
\begin{subequations}
\begin{align}
Q_0&\equiv\int_{-a/2}^{R_c-a/2}\!\!\!\!dx\,\rho_0(x)\\
&=\sum_n\braket{E=0,n,\mathrm{L}|\Gamma|E=0,n,\mathrm{L}}\\
&=n_+-n_-,
\end{align}
\end{subequations}
where $\ket{E=0,n,\mathrm{L}}$ specifies gapless end states localized at the left edge.
{}{Here, we set $\Gamma^2=1$, following the usual convention.}
Thus, $Q_0$ is the difference of the number of gapless left-end states with positive and negative chiralities.
According to the index theorem, this is equivalent to the winding number $w_0=w(0)$\Cite{Sato2011,Gurarie2011}.

Next, we consider $Q_n$ for $n\ge 1$.
$Q_n$ is obtained by a discussion parallel to that of $Q$ (\Sec{sec:sec2B}).
{}{Let us consider the replacement of electric charge with chirality charge, $\rho(x)\to\rho_f(x,z)$, in the relation $Q=-p+p_{\mathrm{cl}}$.
This replacement is valid since chirality charges satisfy the two neutrality conditions, Eqs.~\eqref{eq:lcn_f} and \eqref{eq:cn_tot}, as is the case for electric charge (Eqs.~\eqref{eq:tcn} and \eqref{eq:lcn}).
Thus, we obtain}
\begin{equation}
  F(z)=-\frac{1}{L}\int_{-\infty}^\infty dx\,x\rho_f(x,z) +\frac{1}{a}\int_{-a/2}^{a/2}dr\,r\rho_f(R_c+r,z).\label{eq:eqFz}
\end{equation}
Laurent expansion of the lefthand side gives $Q_n$ as coefficients.
On the other hand, coefficients of the first term in the right hand side give, from \Eq{eq:rho_n},
\begin{subequations}
\begin{align}
  &-\frac{1}{L}\int^\infty_{-\infty}dx\,x\rho_n(x)=-\frac{1}{L}2\Tr[\Gamma H^{-2n}\hat{P}_{\mathrm{occ}}\hat{x}]\\
&=-\frac{2}{L}\sum_{R,\nu}\braket{W_\nu(R)|\hat{x}\Gamma H^{-2n}\hat{P}_{\mathrm{occ}}|W_\nu(R)}\\
&=-\frac{2}{L}\sum_{R,\nu}\braket{W_\nu(R)|(\hat{x}-R)\Gamma H^{-2n}\hat{P}_{\mathrm{occ}}|W_\nu(R)}\label{eq:32c}\\
&=-\frac{2}{a}\sum_\nu\braket{w_\nu(0)|\hat{x}\Gamma H^{-2n}|w_\nu(0)}.
\end{align}
\end{subequations}
The third line follows since $\Gamma H^{-2n}\hat{P}_{\mathrm{occ}}\ket{W_\nu(R)}$ belongs to the subspace spanned by unoccupied states. This point is related to the charge neutrality discussed in the previous subsection.
The forth line follows by identifying $\hat{P}_{\mathrm{occ}}\ket{W_\nu(R)}$ with $\ket{W_\nu(R)}$, and GWFs with Wannier functions, for bulk lattice points.
The coefficients of the second term of the righthand side of \Eq{eq:eqFz} can be rewritten as
\begin{subequations}
  \begin{align}
  &\frac{1}{a}\int^{a/2}_{-a/2}dr\,r\rho_n(R_c+r,z)\\
  &=\frac{2}{a}\sum_\nu\braket{w_\nu(0)|\Delta\hat{r}\Gamma H^{-2n}|w_\nu(0)}.
\end{align}
\end{subequations}
Thus, we obtain
\begin{equation}
Q_n=-\frac{2}{a}\sum_\nu\braket{w_\nu(0)|\hat{R}\Gamma H^{-2n}|w_\nu(0)}.
\label{eq:Qn}
\end{equation}
This expression for chirality charge is analogous to that for electric charge \Eq{eq:eqQ}.
Now the accumulated chirality charges are represented by bulk properties.
Therefore, we can relate $Q_n$ to $w_n$.

Based on the above equation, we can easily show $Q_n=w_n$, after some algebra [see \App{sec:SBBCcalc}].
Thus, SBBC $F(z)=w(z)/z$ has been proved.
{Let us stress again that $\omega_n$ is a bulk quantity while $Q_n$ (Eq.~\eqref{eq:Q_n}) is a surface quantity.
The geometric BBC between $\omega_n$ and $Q_n$ $(n\ge1)$ is an essential origin of SBBC. }

We also note that our derivation gives an intuitive explanation for the reported robustness of SBBC against surface disorders preserving the chiral symmetry\Cite{Tamura2018}. In the summand of \Eq{eq:32c}, deviation of $\ket{W_\nu(R)}$ from Wannier functions in perfect crystal would decay exponentially as $R$ gets away from impurities\Cite{Kohn1973}. Thus, $Q_n$ in the thermodynamic limit is given by $w_n$, since the number of GWFs modified by surface disorders is not of $O(L)$.
In reality, the deviation of $F(z)$ from $w(z)/z$ is shown to be exponentially small as $L\to\infty$ (\App{sec:impuritySBBC}), in agreement with the numerics in Ref.~\cite{Tamura2018}.
Thus, SBBC holds regardless of the details of boundaries.

\section{Conditions for BBC of polarizations}
\label{sec:comment}
In this section, we discuss the similarities and differences between various kinds of polarizations.
First, we comment on \textit{spin density polarization} $\tr[s_i\hat{x}_j\theta(-H)]$\Cite{Chen2018,Thole2016}.
This quantity was investigated in Ref.~\cite{Chen2018} based on GWFs as in the present paper.
However, a \textit{gauge-dependent} wave-number-space expression has been reported, in contrast to the gauge-independent formula of chirality polarizations obtained by us.
Since the similar method is used, it is important to clarify what makes difference between these two quantities.

For simplicity, we consider spin density polarization of one-dimensional systems.
Rewriting the definition, we obtain
\begin{align}
\tr[s_i\hat{x}\theta(-H)]=\sum_{R,\nu}&\braket{W_\nu(R)|s_i(\hat{x}-R)|W_\nu(R)}\notag\\
&+\sum_{R,\nu}R\braket{W_\nu(R)|s_i|W_\nu(R)}.\label{eq:spin_pol}
\end{align}
The authors of Ref.~\cite{Chen2018} claim that the second term vanishes except for ferromagnets and ferrimagnets.
However, we would say this is not the case in general, because the GWFs around the surface might have finite spin expectation value, even for antiferromagnets. We exemplify this case in \App{sec:spinpol} for a spin-orbit coupled antiferromagnetic chain. In addition, {contribution from the surface GWFs is expected to change by an arbitrarily small amount with retaking the basis set of GWFs}, while $\tr[s_i\hat{x}\theta(-H)]$ does not [see \App{sec:spinpol}]. This would be the reason why the first term of \Eq{eq:spin_pol} and its resultant expression in the wave-number space are gauge dependent.
Once contribution from the surface GWFs is properly included, if possible, spin density polarization might have gauge-independent expression, as is the case for orbital magnetic dipole moments\Cite{Ceresoli2006,Souza2008}.

Next, let us reconsider electric polarization.
In this case, $s_i$ in \Eq{eq:spin_pol} is replaced with unity, and then the second term is independent of the basis choice of the GWFs (though origin-dependent).
Thus, we can safely pass to wave-number space with using Wannier functions associated with each lattice point \textit{in the same manner} as GWFs, which corresponds to a specific gauge fixing.
Such a gauge associated with polarization in OBC is sensitive to surface geometry, and indeterminable from the bulk.
However, non-integral part of \Eq{eq:BerryPhaseFormula} is gauge independent, and therefore, replacement of $\ket{W_\nu(R)}$ with $\ket{w_\nu(R)}$ defined by \Eq{eq:Wannier_def} predicts the correct value.
This is why $p_{\mathrm{el}}$ modulo lattice constant is the bulk property.
Note that the same situation holds for spin density polarization in collinear magnets in the absence of spin-orbit coupling, since the discussion for electric polarization holds in each spin sector\Cite{Thole2016,Batista2008}. Thus, \textit{polarization of charges conserved by local symmetries are bulk properties modulo polarization quantum}.

Finally, as for chirality polarizations, the second term of \Eq{eq:spin_pol} with replacing $s_i$ with $\hat{e}_n$ identically vanishes. In addition, the first term includes only the interband component of the position operator, which is gauge invariant.
These properties arise from the local antisymmetry.
Thus, chirality polarizations, that is, \textit{polarizations of charges corresponding to local antisymmetries, are bulk properties}.
This means that components of surface accumulation of Green's function (as many as the number of local antisymmetries) are determined by the bulk.
{Thus, local antisymmetries impose stronger constraints on surface physics than local symmetries.
In other words, BBC by local antisymmetries is stronger than BBC by local symmetries.
This is especially important for time-reversal symmetric superconductivity, where various antisymmetries may intrinsically emerge in combination with crystalline symmetries.}

\section{Formal proof of SBBC}
\label{sec:derivation}
The proof based on GWFs is useful to get an intuitive picture of SBBC, but it relies on some physical assumptions for GWFs.
Although we believe the derivation reasonable enough,
we present an alternative route to prove SBBC for completeness.

The idea is again based on the similarity to electric polarization.
As is well known, change of the surface electric charge is equivalent to the transient current through the bulk region:
\begin{equation}
\frac{d}{dt}\int^{x_c}_{-\infty}dx\,\bar{\rho}(x)=-\bar{j}(x_c).\label{eq:current_pol}
\end{equation}
It is expected that change of the chirality charges can also be characterized by the bulk quantity.
This is indeed true, and below we show
\begin{equation}
  \partial_\lambda [zF(z)]=\partial_\lambda w(z)
  \label{eq:delSBBC}
\end{equation}
for adiabatic change of Hamiltonian specified by a parameter $\lambda$. Along with the fact that SBBC holds for atomic insulators, we will conclude that SBBC generally holds for one-dimensional chiral symmetric systems.
Below, we show the details of the derivation.

Let us consider tight-binding models for clarity of discussion.
In the following, we set $a=1$ for simplicity.
For tight-binding models, $F(z)$ can be concisely rewritten as
\begin{equation}
F_\lambda(z)=\sum_{R\le R_c}\Tr[n(R)\Gamma G_\lambda(z)],
\end{equation}
where $n(R)$ is the charge density operator within a unit cell:
\begin{equation}
n(R)\equiv\sum_\alpha\ket{R,\alpha}\bra{R,\alpha}.
\end{equation}
The internal degrees of freedom $\alpha$ include sublattice degrees of freedom.
{}{Chiral operator $\Gamma$ is assumed to be local, i.e. $[n(R),\Gamma]=0$.}
We investigate adiabatic evolution of $F_\lambda(z)$, in response to a continuous deformation of the Hamiltonian $H_\lambda=H_{\mathrm{atom}}+\lambda(H-H_{\mathrm{atom}})$,
with $G_\lambda(z)\equiv(z-H_\lambda)^{-1}$.
For a while, we consider OBC.

Let us calculate the derivative of $F_\lambda(z)$. By using $\partial_\lambda G_\lambda (z)=G_\lambda(z)\partial_\lambda H_\lambda G_\lambda(z)$,
\begin{subequations}
\begin{align}
\partial_\lambda F_\lambda(z)=&\sum_{R\le R_c}\Tr\bigl[[G_\lambda(z),\,n(R)]\Gamma G_\lambda(z)\partial_\lambda H_\lambda\bigr]\\
&\ -\sum_{R\le R_c}\Tr\bigl[n(R)\Gamma G_\lambda(-z)G_\lambda(z)\partial_\lambda H_\lambda\bigr].
\end{align}
\end{subequations}
The second term is odd in $H_\lambda$, and thus vanishes due to chiral symmetry.
Roughly speaking, the first term includes $[H_\lambda, n(R)]\sim \partial_Rj(R)$, and contribution from only around $R=R_c$ remains after integration over $R$.
Using this property we will pass to PBC from OBC.
We show a formal derivation in the following discussion.
Let us introduce Peierls phase into Hamiltonian by
\begin{widetext}
\begin{equation}
H_\lambda[A]\equiv \sum_{R_1,R_2,\alpha,\beta}\ket{R_1,\alpha}\braket{R_1,\alpha|H_\lambda|R_2,\beta}e^{iA_{R_1,R_2}}\bra{R_2,\beta}.
\end{equation}
\end{widetext}
Here, the link variable $A_{R_1,R_2}$ is defined to be antisymmetric in accordance with Hermitian properties of the Hamiltonian, and we take $\Set{A_{R_1,R_2}|\ R_1>R_2}$ as a set of independent variables. We also define $G_\lambda[A]\equiv (z-H_\lambda[A])^{-1}$.
Following the procedure to derive equation of continuity, we consider an infinitesimal local gauge transformation
\begin{equation}
U_\epsilon(R)\equiv \exp(i\epsilon n(R)),\quad\epsilon\ll 1.
\end{equation}
Then, since
\begin{equation}
  U_\epsilon(R)G_\lambda U_\epsilon(R)^\dagger=G_\lambda[\epsilon A(R)]
\end{equation}
with $A(R)_{R_1,R_2}=\delta(R_1,R)-\delta(R,R_2)$, we obtain
\begin{subequations}
\begin{align}
[G_\lambda,n(R)]&=\frac{U_\epsilon(R)G_\lambda U_\epsilon(R)^\dagger-G_\lambda}{-i\epsilon}\\
&=i\sum_{R_1>R_2}\frac{\partial G_\lambda}{\partial A_{R_1,R_2}}A(R)_{R_1,R_2}.
\end{align}
\end{subequations}
Thus, we obtain
\begin{equation}
\partial_\lambda F_\lambda(z)=i\sum_{R\le R_c,\,R'}\Tr\left[\frac{\partial G_\lambda}{\partial A_{R',R}}\Gamma G_\lambda(z)\partial_\lambda H_\lambda\right],
\end{equation}
where we used the antisymmetry ${\partial G_\lambda}/{\partial A_{R_1,R_2}}=-{\partial G_\lambda}/{\partial A_{R_2,R_1}}$.
Note that contribution from the region $R'\le R_c$ vanishes again due to the antisymmetry.
Furthermore, we retake the set of independent variables from $(R,R')$ to $(\Delta R\equiv R'-R,\ R')$. Then, we reach an important expression,
\begin{equation}
\partial_\lambda F_\lambda(z)=i\sum_{\substack{0< \Delta R;\\R_c\le R'<R_c+\Delta R}}\Tr\left[\frac{\partial G_\lambda}{\partial A_{R',R'-\Delta R}}\Gamma G_\lambda(z)\partial_\lambda H_\lambda\right].\label{eq:delF}
\end{equation}
Note that $\partial G_\lambda/\partial A_{R',R'-\Delta R}$ involves $\partial H_\lambda/\partial A_{R',R'-\Delta R}$, which is proportional to the hopping amplitude between the unit cells specified by $R'$ and $R'-\Delta R$. Therefore, the summation for $\Delta R$ has an intrinsic cutoff determined by the range of the transfer integral, and we have only to consider $R'$ and $R'-\Delta R$ as bulk lattice points.
Based on this observation, we find that $\partial_\lambda F_\lambda(z)$ is determined by the bulk matrix elements of the Green's functions such as
\begin{subequations}
\begin{align}
&\braket{R_1,\alpha|G_\lambda(z)\Gamma G_\lambda(z)\partial_\lambda H_\lambda G_\lambda(z)|R_2,\beta}\\
&=\braket{R_1,\alpha|G_\lambda(z)\Gamma \partial_\lambda(G_\lambda(z)|R_2,\beta}).\label{eq:replace}
\end{align}
\end{subequations}
It is known that Green's function of gapped systems damps exponentially in space.
For this reason, the effect of the surfaces is negligible in the bulk matrix elements, and the Green's function and Hamiltonian can be replaced with those in PBC.
This replacement remains valid even when the system is gapless, as long as $\Im z\neq0$ is concerned.
This is simply because $G(z)$ for $\Im z\neq0$ decays in space even for metals and nodal superconductors, owing to the introduced lifetime $1/|\Im z|$ of Bloch waves [see \Sec{sec:decayA} for a formal derivation].

In the following, we pass to PBC and the quantities $G$ and $H$ represent Green's function and Hamiltonian for PBC, respectively.
Taking translational symmetry into account, we obtain
\begin{widetext}
\begin{subequations}
\begin{align}
\partial_\lambda F_\lambda(z)&=\frac{i}{N}\sum_{\Delta R>0}\Delta R\sum_R\Tr\left[\frac{\partial G_\lambda}{\partial A_{R+\Delta R,R}}\Gamma G_\lambda(z)\partial_\lambda H_\lambda\right]\\
&=\sum_{k,\alpha,\beta}\frac{i}{N}\sum_{\Delta R>0}\Delta R\sum_R\braket{k,\alpha|\frac{\partial H_\lambda}{\partial A_{R+\Delta R,R}}|k,\beta}\left(G_{k\lambda}(z)\Gamma G_{k\lambda}(z)\partial_\lambda H_{k\lambda}G_{k\lambda}(z)\right)_{\beta\alpha}.
\end{align}
\end{subequations}
Here, $\ket{k,\alpha}$ is the Fourier transform of $\ket{R,\alpha}$,
\begin{equation}
\ket{k,\alpha}\equiv\frac{1}{\sqrt{N}}\sum_Re^{ikR}\ket{R,\alpha},
\end{equation}
while $H_k$ and $G_k$ are the matrices
\begin{equation}
(H_k)_{\alpha\beta}=\braket{k,\alpha|H|k,\beta},\quad G_k\equiv(z-H_k)^{-1}.
\end{equation}
Let us write the above-mentioned cutoff of $\Delta R$ explicitly as $l_c$.
Then,
\begin{subequations}
\begin{align}
i\sum_{0<\Delta R<l_c}\Delta R\sum_R\braket{k,\alpha|\frac{\partial H_\lambda}{\partial A_{R+\Delta R,R}}|k,\beta}
&=\frac{-1}{N}\sum_{|\Delta R|<l_c}\Delta R\sum_R\braket{R+\Delta R,\alpha|{H_\lambda}|R,\beta}e^{-ik\Delta R}\\
&=\frac{1}{iN}\partial_k\sum_{|\Delta R|<l_c}\sum_R\braket{R+\Delta R,\alpha|{H_\lambda}|R,\beta}e^{-ik\Delta R}\\
&=-i\partial_k(H_{k\lambda})_{\alpha\beta}.
\end{align}
\end{subequations}
Finally, we obtain
\begin{equation}
\partial_\lambda F_\lambda(z)=\int\frac{dk}{2\pi i}\tr\left[\partial_kH_{k\lambda}\,G_{k\lambda}(z)\Gamma G_{k\lambda}(z)\partial_\lambda H_{k\lambda}\,G_{k\lambda}(z)\right].
\label{eq:SBBCtemp}
\end{equation}
\end{widetext}
In \App{sec:derivation_of_SBBC}, we show that the righthand side is equivalent to $\partial_\lambda w(z)/z$ after somewhat technical calculation.
Thus, $\partial_\lambda zF_\lambda(z)=\partial_\lambda w_\lambda(z)$ holds.
This relation is valid for $\Im z\neq0$ even right on the topological phase transition points.

Next we show the equivalence of $zF_0(z)$ with $w_0(z)$.
Note that we can always choose the chiral operator so as to satisfy
\begin{equation}
\braket{R_1,(\bar{\alpha},i)|\Gamma|R_2,(\bar{\beta},j)}=\delta_{R_1,R_2}\delta_{\bar{\alpha}\bar{\beta}}(\sigma_x)_{ij},
\end{equation}
by retaking basis of internal degrees of freedom as $\alpha=(\bar{\alpha},i)$.
{}{[When we do not adopt $\Gamma^2=1$, $\delta_{\bar{\alpha}\bar{\beta}}$ should be replaced with some unitary matrix.]}
Let us consider an atomic insulator
\begin{equation}
\braket{R_1,(\bar{\alpha},i)|H_{\mathrm{atom}}|R_2,(\bar{\beta},j)}=\delta_{R_1,R_2}\delta_{\bar{\alpha}\bar{\beta}}(\sigma_z)_{ij}.
\end{equation}
We can readily calculate $zF_0(z)$, as
\begin{equation}
zF_0(z)\propto\tr\left[\sigma_x\begin{pmatrix}1/(z-1)&0\\0&1/(z+1)\end{pmatrix}\right]=0.
\end{equation}
On the other hand, $w_0(z)$ also vanishes since $\partial_k H_{k0}=0$ for atomic insulators.
Thus, $zF_0(z)=w_0(z)$ holds.

From the above relation and \Eq{eq:delSBBC}, we now know that $zF(z)=w(z)$ holds for $\Im z\neq0$.
This equality can be extended to $z\sim0$ by analytic continuation.
Let us define $f(z)\equiv zF(z)=w(z)$ as a regular function in $\Im z\neq0$.
When the system in PBC is gapful, both $zF(z)$ and $w(z)$ are analytic around $z\sim 0$ as well as for $\Im z\neq0$,
as evident in the expansion by energy eigenstates.
Hence, they are the analytic continuation of $f(z)$ into the region $z\sim 0$.
It immediately follows that $zF(z)=w(z)$ holds because of the uniqueness of analytic continuation.
Thus, SBBC has been proved, including a proof of topological BBC\Cite{Essin2011,Sato2011,Gurarie2011} as a special case.


\section{Conclusion}
\label{sec:conclusion}
In this paper, we introduced the notion of chirality polarizations and showed how surface chirality charges are characterized by the bulk.
Thus, geometric bulk-boundary correspondence for chirality charges has been established.
Our discussion, at the same time, proves the spectral bulk-boundary correspondence, which is the generalization of the bulk-boundary correspondence into complex frequencies.
We showed a {physically transparent proof} via Wannier functions and a formal proof by deformation of the Hamiltonian.
We have also clarified the importance of the gauge-invariance as well as the charge neutrality conditions to obtain meaningful expression in wave-number space, by comparing spin-density, electric, and chirality polarizations.
Our work, together with \Ref{Tamura2018}, has revealed that the surface physics is dominated to a large extent by the bulk, in systems with local antisymmetries such as superconductors.

Finally, we note that our result is limited to one-dimensional case as well as the case for clear surfaces of higher-dimensional systems (by fixing a k-point in the surface Brillouin zone). It is an interesting question whether similar correspondence holds for higher-dimensional winding numbers, and how it is related to higher-rank multipole moments.

\begin{acknowledgments}
This work was inspired by a fruitful discussion with S. Tamura.
The authors are also grateful to Y. Tanaka for helpful discussions about odd-frequency pairings.
This work was supported by Grant-in Aid for Scientific Research on Innovative Areas ``J-Physics'' (No.~15H05884) and ``Topological Materials Science'' (No.~18H04225) from JSPS of Japan, and by JSPS KAKENHI Grants No. 15K05164, No. 15H05745, No. 17J10588, No.~18H01178, and No.~18H05227.
\end{acknowledgments}


%

\appendix


\begin{table*}[pt]
\centering
\begin{tabular}{l@{\hspace{0.5cm}}l}
\hline\hline
quantity&definition\\\hline
GWFs in OBC&$\ket{W_\nu(R)}$\\
Wannier functions in PBC &$ \ket{w_\nu(R)}=\frac{1}{\sqrt{N}}\sum_ke^{-ikR}\ket{\psi_\nu(k)}$\\
energy eigenstates of $H$ in PBC &$\ket{\psi_\nu(k)}=\ket{\psi_\nu(k+2\pi)}$\\
periodic part of Bloch wave (for polarization)& $\ket{u_\nu(k)}=e^{-ik\hat{x}}\ket{\psi_\nu(k)}=e^{-i2\pi\hat{x}}\ket{u_\nu(k+2\pi)}\qquad$\\
periodic part of Bloch wave ($k$ periodic) &$\ket{U_\nu(k)}=e^{-ik\hat{R}}\ket{\psi_\nu(k)}=\ket{U_\nu(k+2\pi)}$\\
position operator&$\hat{x}=\sum_s\int dx\,x\ket{x,s}\bra{x,s}$\\
lattice point operator&$\hat{R}=\sum_{s,R}R\int^{a/2}_{-a/2} dr\ket{R+r,s}\bra{R+r,s}$\\
\hline\hline
\end{tabular}
\caption{Summary of notations for Wannier functions.}
\label{tab:notations_WFs}
\end{table*}

\section{Summary of notations}
\label{sec:notations}
We first summarize the notations used in the main text (so this section might be skipped).
\subsection{Notations for electric polarization}
Table~\ref{tab:notations_WFs} shows the notations for Wannier functions.
GWFs are defined in OBC and written as $\ket{W_\nu(R)}$.
On the other hand, Wannier functions are defined in PBC and written as $\ket{w_\nu(R)}$ and are the Fourier transformation of the energy eigenstates (or the Bloch waves) are $\ket{\psi_\nu(k)}$, whose gauge is chosen to be periodic in $k$.
The periodic (in terms of lattice translation) part of the Bloch waves is defined in two ways. The first one is $\ket{u_\nu(k)}$, which is used for the calculation of electric polarization. The other is $\ket{U_\nu(k)}$, which is used for the calculation of the microscopic excess charge. Only the latter one is periodic in $k$. Correspondingly, there are two kinds of ``postition operators''. First one is the genuine position operator $\hat{x}$, while the other one $\hat{R}$ captures only the lattice points of each position eigenstate $\ket{x,s}$.
Therefore, the operator $\hat{R}$ is dependent on the choice of the unit cell, which is specified through the surface termination.

Table~\ref{tab:notations_lc} shows the notations for local electric charge density.
The microscopic local charge density is written as $\rho(x)$, which is divided into two parts: electronic and ionic contribution.
Electronic contribution is given by the sum of the contribution from all the GWFs.
On the other hand, ions are assumed to be classical point charges and their contribution comes in the form of delta functions.
The coarse-grained local charge density is defined as the microscopic charge averaged over a unit cell, and used for the definition of surface bound charge. Importantly, $\bar{\rho}(x)$ vanishes deep inside the sample.

\begin{table*}[hbt]
\centering
\begin{tabular}{l@{\hspace{0.5cm}}l}
\hline\hline
quantity&definition\\\hline
microscopic local charge density& $\rho(x)=\rho_{\mathrm{el}}(x)+\rho_{\mathrm{ion}}(x)$\\
electronic contribution to $\rho(x)$&$\rho_{\mathrm{el}}(x)=\sum_{R,\nu}|\braket{x|W_\nu(R)}|^2$\\
ionic contribution to $\rho(x)$&$\rho_{\mathrm{ion}}(x)=\sum_{R,i}Z_i\,\delta(x-R-u_i(R))$\\
coarse-grained local charge density&$\bar{\rho}(x)=\frac{1}{a}\int^{a/2}_{-a/2}\rho(x+r)$\\
\hline\hline
\end{tabular}
\caption{Summary of notations for local electric charge density.}
\label{tab:notations_lc}
\end{table*}

Table~\ref{tab:notations_pol} shows the notations for electric polarization.
Electric polarization is defined through $\rho(x)$, and is divided into electronic and ionic contributions.
Electronic contribution $p_{\mathrm{el}}$ is calculated by the Wannier centers, which is equivalent to the well-known Berry phase formula (Eq.~(5b) in the main text).
We also add ionic contribution $p_{\mathrm{ion}}$ to the table, although it is not defined in the main text.
It is just written as the shift of point charges from the unit cell centers.

\begin{table*}[tbh]
\centering
\begin{tabular}{l@{\hspace{0.5cm}}l}
\hline\hline
quantity&definition\\\hline
electric polarization&$p=\frac{1}{L}\int^\infty_{-\infty}dx\,x\rho(x)=p_{\mathrm{el}}+p_{\mathrm{ion}}$\\
electronic contribution to $p$&$p_{\mathrm{el}}=\frac{1}{a}\sum_\nu\braket{w_\nu(0)|\hat{x}|w_\nu(0)}$\\
ionic contribution to $p$&$p_{\mathrm{ion}}=\frac{1}{L}\sum_{R,i}Z_i\,u_i(R)$\\
\hline\hline
\end{tabular}
\caption{Summary of notations for electric polarization.}
\label{tab:notations_pol}
\end{table*}

Table~\ref{tab:notations_sc} shows the notations for surface bound charge.
Electric polarization $-p$ is equivalent to the coarse-grained surface charge $\sigma_B$ bound at the left end.
Physically, $\sigma_B$ is divided into two parts. The first one is the microscopic excess charge $Q$, which is the integration of microscopic local charge over sufficiently large number of unit cells counted from the left end.
The other one is the classical polarization $p_{\mathrm{cl}}$, which is the charge asymmetry within the bulk unit cell.

\begin{table*}[tbh]
\centering
\begin{tabular}{l@{\hspace{1cm}}l}
\hline\hline
quantity&definition\\\hline
coarse-grained surface chrage&$\sigma_B=\int^{x_c}_{-\infty}dx\,\bar{\rho}(x)=-p=-p_{\mathrm{cl}}+Q\qquad$\\
microscopic excess charge&$ Q=\int^{R_c-a/2}_{-a/2}\! dx\,\rho(x)=-\frac{1}{a}\sum_\nu\braket{w_\nu(0)|\hat{R}|w_\nu(0)}$\\
classical polarization&$ p_{\mathrm{cl}}=\frac{1}{a}\int^{a/2}_{-a/2}dr\,r\rho(R_c+r)=\frac{1}{a}\sum_\nu\braket{w_\nu(0)|(\hat{x}-\hat{R})|w_\nu(0)}{+\frac{1}{L}\sum_{R,i}Z_i\,u_i(R)}$\\
\hline\hline
\end{tabular}
\caption{Summary of notations for surface electric charge.}
\label{tab:notations_sc}
\end{table*}

\subsection{Notations for chirality polarizations}
Table~\ref{tab:notations_Hami} shows the notations for Green's functions.
The Green's function is defined by the usual way via Hamiltonian $H$.
Projection operators are defined for the space of negative-, positive-, and zero-energy states.
 In PBC, wave-number space expression of the Green's function is defined via the Bloch Hamiltonian $H_k$, which is periodic in $k$ for this definition.

\begin{table*}[htb]
\centering
\begin{tabular}{lc}
\hline\hline
quantity&definition\\\hline
Hamiltonian &$H$\\
Green's function &$G=(z-H)^{-1}$\\
projection operator onto zero-energy states&$\hat{P}_0$\\
projection operator onto negative-energy states&$\hat{P}_{\mathrm{occ}}$\\
projection operator onto positive-energy states&$\hat{P}_{\mathrm{unocc}}$\\
Bloch Hamiltonian&$H_k=e^{-ik\hat{R}}He^{ik\hat{R}}$\\
Green's function in wave-number space &$G_k(z)=(z-H_k)^{-1}$\\
\hline\hline
\end{tabular}
\caption{Summary of notations for Green's function.}
\label{tab:notations_Hami}
\end{table*}

Table~\ref{tab:notations_w} shows the notations for the generalized winding number.
The generalized winding number $w(z)$ is defined by replacing $H_k$ with $G_k$ in the expression of the winding number $w(0)$.
We call each coefficient of the Taylor expansion of $w(z)$ as $w_n$.

\begin{table*}[htb]
\centering
\begin{tabular}{l@{\hspace{0.5cm}}l}
\hline\hline
quantity&definition\\\hline
generalized winding number&$w(z)\equiv-\int\frac{dk}{4\pi i}\tr\left[\Gamma G_k(z)\partial_k G_k^{-1}(z)\right]=z\sum_n\,w_nz^{2n-1}\qquad$\\
$n$-th coefficient of $w(z)$ &$w_n=-\int\frac{dk}{4\pi i}\tr\left[\Gamma H_k^{-(2n+1)}\partial_k H_k\right]$\\
\hline\hline
\end{tabular}
\caption{Summary of notations for the generalized winding number.}
\label{tab:notations_w}
\end{table*}

Table~\ref{tab:notations_cp} shows the notations for the local chirality charges.
The $0$-th chirality charge is determined by the zero-enegy states, while $n$-th ones are determined by the negative-energy states.
They are expressed as the coefficient of Laurent expansion of a condensed notation $\rho_f(x,z)$.
In analogy with electric polarization, $\bar{\rho}_f(x,z)$ specifies the coarse-grained local chirality charges, which vanishes deep inside the sample. We discuss Laurent expansion of $\rho_f(x,z)$ and derive expression of $\rho_n(x)$ in the next section.

\begin{table*}[htb]
\centering
\begin{tabular}{l@{\hspace{1cm}}l}
\hline\hline
quantity&definition\\\hline
$0$-th local chirality charge&$\rho_0(x)=\sum_s\braket{x,s|\Gamma \hat{P}_0|x,s}$\\
$n$-th local chirality charge &$\rho_n(x)=\sum_s2\braket{x,s|\Gamma H^{-2n}\hat{P}_{\mathrm{occ}}|x,s}\quad(n\ge1)$\\
condensed notation for local chirality charges&$\rho_f(x,z)=\sum_s\braket{x,s|\Gamma G(z)|x,s}=\sum_n\rho_n(x)z^{2n-1}$\qquad\\
coarse-grained local chirality charges&$\bar{\rho}_f(x,z)=\frac{1}{a}\int^{a/2}_{-a/2}dr\,\rho_f(x+r,z)$\\
\hline\hline
\end{tabular}
\caption{Summary of notations for the local chirality charges.}
\label{tab:notations_cp}
\end{table*}

\begin{table*}[th]
\centering
\begin{tabular}{l@{\hspace{1cm}}l}
\hline\hline
quantity&definition\\\hline
$0$-th chirality polarization&$Q_0=\int_{-a/2}^{R_c-a/2}\! dx\,\rho_0(x)=n_+-n_-$\\
$n$-th chirality polarization&$Q_n=\int_{-a/2}^{R_c-a/2}\! dx\,\rho_n(x)$\\
Surface-accumulated Green's function &$F(z)=\int_{-a/2}^{R_c-a/2} dx\,\rho_f(x,z)=\sum_nQ_nz^{2n-1}$\qquad\\
\hline\hline
\end{tabular}
\caption{Summary of notations for the chirality polarizations (or the surface chirality charge).}
\label{tab:notations_cp2}
\end{table*}

Table~\ref{tab:notations_cp2} shows the notations for chirality polarizations, which correspond to the ``microscopic excess charge $Q$'' for the local chirality charges. They can be summarized into a condensed notation $F(z)$, which is nothing but the component of the surface-accumulated Green's function proportional to the chiral operator. Expression of $Q_n$ in terms of GWFs is discussed in the next section.

\renewcommand{\arraystretch}{1}

\section{Derivation of $Q_n=w_n$}
\label{sec:SBBCcalc}
In this section, we show some calculations for the derivation of $Q_n=w_n$ skipped in the main text.

\subsection{Laurent expansion of $\rho_f(x,z)$}
Here, we derive expression of local chirality charges listed in \Table{tab:notations_cp}.
Let us expand $\rho_f(x,z)$ by energy eigenstates in OBC:
\begin{equation}
\rho_f(x,z)=\sum_{E,s}\frac{\braket{E|x,s}\braket{x,s|\Gamma|E}}{z-E}.
\end{equation}
Thus, in the Laurent expansion $\rho_f(x,z)=\sum_n\rho_nz^{2n-1}$, contribution to $1/z$ comes only from the zero-energy states.
Thus,
\begin{equation}
\rho_0(x)=\sum_s\braket{x,s|\Gamma\hat{P}_0|x,s}.
\end{equation}
The coefficient of $z^{2n-1}$ is given by
\begin{equation}
\rho_n(x)=\sum_{E,s}\oint_{C_\epsilon}\frac{dz}{2\pi i}\frac{1}{z^{2n}}\frac{\braket{E|x,s}\braket{x,s|\Gamma|E}}{z-E}\quad(n\ge1),
\end{equation}
with $C_\epsilon$ a small loop around $z=0$ whose diameter is smaller than the bulk gap.
Contribution from zero-energy states vanishes. Thus,
\begin{subequations}
\begin{align}
\rho_n(x)&=\sum_{E\neq0,s}\frac{1}{E^{2n}}{\braket{E|x,s}\braket{x,s|\Gamma|E}}\\
&=2\sum_s\braket{x,s|\Gamma H^{-2n}\hat{P}_{\mathrm{occ}}|x,s}\quad(n\ge1),
\end{align}
\end{subequations}
since contribution from positive- and negative-energy states are equivalent.

\subsection{Derivation of $Q_n=w_n$}
Let us complete the derivation of $Q_n=w_n$.
Due to the locality of Wannier functions, we can view $\hat{R}$ in \Eq{eq:Qn} as not of $O(L)$.
Thus, the equality
\begin{equation}
\hat{R}\ket{w_\nu(0)}=\frac{-i}{\sqrt{N}}\sum_k\left(\ket{\partial_k\psi_\nu(k)}-e^{ik\hat{R}}\ket{\partial_kU_\nu(k)}\right)
\end{equation}
is justified. The first term vanishes in the gauge satisfying $\ket{\psi_\nu(k+2\pi)}=\ket{\psi_\nu(k)}$, where $\ket{U_\nu(k+2\pi)}=\ket{U_\nu(k)}$ also holds.
Hence, we obtain
\begin{widetext}
\begin{subequations}
\begin{align}
&Q_n=\frac{2i}{L}\sum_{k,\nu:\mathrm{occ}}\braket{\partial_kU_\nu(k)|\Gamma H_k^{-2n}|U_\nu(k)}\\
&=\frac{i}{L}\sum_{k,\rho:\mathrm{all}}\braket{\partial_kU_\rho(k)|\Gamma H_k^{-2n}|U_\rho(k)}\\
&=\frac{-i}{2L}\sum_{k}\tr\left[\Bigl[H_k,\,\sum_{\rho:\mathrm{all}}\ket{U_\rho(k)}\bra{\partial_kU_\rho(k)}\Bigr]H_k^{-(2n+1)}\Gamma\right].
\end{align}
\end{subequations}
Here, $\rho$ runs over both occupied states and unoccupied states, while $\nu$ runs over only occupied states.
Using the equality
\begin{equation}
\Bigl[H_k,\,\sum_{\rho:\mathrm{all}}\ket{U_\rho(k)}\bra{\partial_kU_\rho(k)}\Bigr]=\partial_k H_k-\sum_{\rho,\rho':\mathrm{all}}\ket{U_\rho(k)}\partial_k\bigl(\braket{U_\rho(k)|H_k|U_{\rho'}(k)}\bigr)\bra{U_{\rho'}(k)},
\end{equation}
we obtain
\begin{equation}
Q_n=\int\frac{dk}{4\pi i}\tr\left[(\partial_kH_k)H_k^{-(2n+1)}\Gamma\right]+\frac{i}{2L}\sum_{k;\,\rho,\rho':\mathrm{all}}\partial_k\bigl(\braket{U_\rho(k)|H_k|U_{\rho'}(k)}\bigr)\,\braket{U_{\rho'}(k)|H_k^{-(2n+1)}\Gamma|U_\rho(k)}.
\end{equation}
\end{widetext}
In the summation, the two matrix elements can not be finite at the same time.
Finally, we obtain
\begin{equation}
Q_n=-\int\frac{dk}{4\pi i}\tr[\Gamma H_k^{-(2n+1)}(\partial_k H_k)]=w_n.
\end{equation}

The discussion above is straightforwardly extended to the cases where several bands are entangled.
To do this, we just reinterpret $\ket{w_\nu(R)}$ as the multiband Wannier functions rather than the Wannier functions deifned in~\Table{tab:notations_Hami}, by replacing $\ket{\psi_\nu(k)}$ and $\ket{U_\nu(k)}$ with $\ket{\tilde{\psi}_\nu(k)}$ and $\ket{\tilde{U}_\nu(k)}$, respectively.
The quantities with tilde are defined by
\begin{subequations}
\label{eq:WannierMultiband}
\begin{gather}
\ket{\tilde{\psi}_\nu(k)}\equiv e^{ik\hat{R}}\ket{\tilde{U}_\nu(k)},\\
\ket{\tilde{U}_\nu(k)}\equiv \sum_{\mu:\mathrm{occ}}\ket{U_\mu(k)}V_{\mu\nu}(k),\label{eq:multibandWannier}
\end{gather}
\end{subequations}
where $V_{\mu\nu}(k)$ is a unitary matrix in the space of occupied states, making $\ket{\tilde{U}_\nu(k)}$ a smooth function of $k$.
Specifically, $\ket{\tilde{U}_\nu(k)}$ is obtained in the twisted parallel-transport gauge\Cite{Marzari1997,Vanderbilt2018}.
Thus, $Q_n=w_n$ holds in general situations.

\section{Gauge dependence of surface GWFs}
\label{sec:spinpol}
In this section, we discuss spin expectation values of GWFs by considering a specific example.
As a consequence, we will show a counterexample of a basic assumption in \Ref{Chen2018}.
Furthermore, it is shown that spin expectation values of the surface GWFs can change by an arbitrary small amount with retaking the basis set of GWFs.

Let us consider an spin-orbit coupled antiferromagnetic chain described by the Hamiltonian
\begin{widetext}
\begin{equation}
\hat{H}=\sum_{R=0}^{L-1}\bm{c}_{{R}}^\dagger(-h_{\mathrm{MF}}\sigma_zs_z-\zeta\alpha\sigma_ys_x)\bm{c}_{R}+\sum_{R=0}^{L-2}\bigl\{\bm{c}_R^\dagger(\alpha s_x\sigma_-)\bm{c}_{R+1}+\mathrm{H. c.}\bigr\},
\end{equation}
with $\bm{c}_R^\dagger=(c_{A\uparrow}^\dagger,c_{A\downarrow}^\dagger,c_{B\uparrow}^\dagger,c_{B\downarrow}^\dagger)$.
Here, Pauli matrices $s_\mu$ and $\sigma_\mu$ specify spin and sublattice degrees of freedom, with $\sigma_-\equiv (\sigma_x-i\sigma_y)/2$.
When PBC is applied, corresponding Hamiltonian is written in the wave-number space as
\begin{subequations}
\begin{gather}
\hat{H}=\sum_{k}\bm{c}_{{k}}^\dagger(-h_{\mathrm{MF}}\sigma_zs_z-\zeta\alpha\sigma_ys_x+\alpha s_x\sigma_y\cos k-\alpha s_x\sigma_x\sin k)\bm{c}_{k},\\
\bm{c}^\dagger_k=\frac{1}{\sqrt{N}}\sum_Re^{ikR}\bm{c}_R^\dagger.
\end{gather}
\end{subequations}
Let us consider the case of half-filling, where two electrons reside in a (magnetic) unit cell on average.
When the molecular field $h_{\mathrm{MF}}$ is sufficiently large, occupied electron states are the eigenstates of $\sigma_zs_z$ with the eigenvalue unity, since $[H,\sigma_zs_z]=0$.
Thus, the space of occupied electron states is spanned by all the eigenstates of the reduced Hamiltonian
\begin{equation}
\hat{H}_+=\sum_k(c^\dagger_{k\,A\uparrow},\,c^\dagger_{k\,B\downarrow})\bigl\{-h_{\mathrm{MF}}+\alpha(\cos k-\zeta)\tau_y-\alpha\sin k\tau_x\bigr\}\begin{pmatrix}c_{k\,A\uparrow}\\ \,c_{k\,B\downarrow}\end{pmatrix}.
\end{equation}
\end{widetext}
This is the SSH model with the pseudo-spin $\tau_\mu$.
The topological phase is obtained for $|\zeta|<1$, ensuring single localized state at each end.

\begin{figure*}
\centering
\includegraphics[width=12cm]{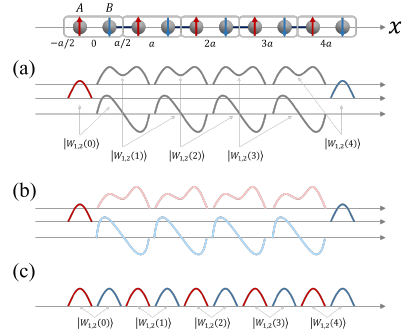}
\caption{(a-c) An antiferromagnetic chain with $L=5a$ and several possible sets of its GWFs. In Fig.~(b), GWFs are assigned to each lattice point in the same way as Fig.~(a). GWFs with positive spin expectation values are highlighted by red and pink, while those with negative spin expectation values are highlighted by blue and light blue. GWFs colored by gray have vanishing spin expectation value.}
\label{fig:SSH}
\end{figure*}

To be specific, we concentrate on the topological limit $\zeta=0$.
Eigenstates of $\hat{H}_+$ in OBC is easily obtained, as shown in \Fig{fig:SSH}(a) for $L=5a$.
They consists of two end states along with the bonding and antibonding orbitals for each inter-cellular bond.
At the same time, they are a possible choice of the set of GWFs, since
\begin{enumerate}
\item They are localized around each unit cell in the sense that $\braket{x|W_\nu(R)}$ decays exponentially as $|x-R|\to\infty$.
\item They coincide with the Wannier functions in PBC $\ket{w_\nu(R)}$, for $R$ in the bulk.
\end{enumerate}
Thus, $\sum_\nu\braket{W_\nu(R)|s_z|W_\nu(R)}$ takes value $\pm1$ for $R=0$ and $L-1$, respectively, while vanishes elsewhere.
This is a counterexample of the claim by the authors of \Ref{Chen2018}.

To go further, let us discuss the basis-set dependence of the spin expectation values of surface GWFs.
It seems that ``contribution from the surface GWFs is expected to change by an arbitrary small amount with retaking the basis set of GWFs'', as quoted from the main text.
Although we don't have an explicit example illustrating this point for GWFs of isolated bands, such a situation \textit{does} occur, for {GWFs for composite bands}.
They are defined to satisfy the conditions $(a)$ and
\begin{enumerate}
\setcounter{enumi}{2}
\item They coincide with the multiband Wannier functions in PBC [see \Eq{eq:WannierMultiband}], for $R$ in the bulk.
\end{enumerate}
Now, let us consider another set of GWFs shown in \Fig{fig:SSH}(b).
We slightly remix the GWFs given in \Fig{fig:SSH}(a) to have finite (but small) spin expectation values.
Then, $|\sum_\nu\braket{W_\nu(R)|s_z|W_\nu(R)}|$ slightly decreases from unity for $R=0$ and $L-1$, while still vanishes elsewhere.
Thus, contribution of the surface GWFs to the term $\sum_{R,\nu} R\braket{W_\nu(R)|s_z|W_\nu(R)}/L$ depends on the gauge choice by $O(1)$.

Another illustrative example is obtained by considering the projected position operator.
Wannier functions are sometimes taken to be its eigenstates.
In our model,
\begin{widetext}
\begin{equation}
\hat{P}_{\mathrm{occ}}\hat{x}\hat{P}_{\mathrm{occ}}=\sum_{R}\left(R-\frac{a}{4}\right)\ket{R,A,\uparrow}\bra{R,A,\uparrow}+\left(R+\frac{a}{4}\right)\ket{R,B,\downarrow}\bra{R,B,\downarrow}
\end{equation}
\end{widetext}
holds, and thus corresponding GWFs are given by \Fig{fig:SSH}(c).
In this case, $\sum_\nu\braket{W_\nu(R)|s_z|W_\nu(R)}$ vanishes for arbitrary $R$.
It should also be noticed that the bulk Wannier functions corresponding to \Fig{fig:SSH}(a) and (b) are obtained by unitary transformations of the form \eqref{eq:multibandWannier}, from those in \Fig{fig:SSH} (c).
This clearly illustrates that gauge-dependence of the wave-number space expression of the spin density polarization is compensated by the contribution from surface GWFs.

In this way, spin expectation values of surface GWFs are highly dependent on the basis-set choice, in contrast to electric and chirarity charges, which are basis independent.
Although we here considered a model with end states, we believe that such a situation occurs quite generally in spin-orbit coupled antiferromagnets.

%

\section{Details of the formal derivation of SBBC}
\label{sec:derivation_of_SBBC}
\subsection{Evaluation of \Eq{eq:SBBCtemp}}
Here, we simplify the righthand side of \Eq{eq:SBBCtemp} and show equivalence of $\partial_\lambda zF_\lambda(z)$ with $\partial_\lambda w_\lambda(z)$.
By multiplying $z$ in each side of the equation,
\begin{widetext}
\begin{subequations}
\begin{align}
\partial_\lambda zF_\lambda(z)&=z\int\frac{dk}{2\pi i}\tr\left[\partial_kH_{k\lambda}\,\Gamma \frac{1}{z^2-H_{k\lambda}^2}\partial_\lambda H_{k\lambda}\,\frac{z+H_{k\lambda}}{z^2-H^2_{k\lambda}}\right]\\
&=\frac{\epsilon_{ij}}{2}\int\frac{dk}{2\pi i}\tr\left[\Gamma \frac{z}{z^2-H_{k\lambda}^2}\partial_i H_{k\lambda}\,\frac{z}{z^2-H^2_{k\lambda}}\partial_jH_{k\lambda}\right],
\end{align}
\end{subequations}
where $\epsilon_{ij}$ is an antisymmetric tensor satisfying $\epsilon_{\lambda k}=1$.
Using the identities
\begin{subequations}
\begin{gather}
\tr\left[\Gamma\frac{H_{k\lambda}}{z^2-H_{k\lambda}^2}\partial_iH_{k\lambda}\frac{z}{z^2-H_{k\lambda}^2}\partial_jH_{k\lambda}\right]=0,\\
\epsilon_{ij}\tr\left[\Gamma\frac{H_{k\lambda}}{z^2-H_{k\lambda}^2}\partial_iH_{k\lambda}\frac{H_{k\lambda}}{z^2-H_{k\lambda}^2}\partial_jH_{k\lambda}\right]=0,
\end{gather}
\end{subequations}
we obtain
\begin{subequations}
\begin{align}
\partial_\lambda zF_\lambda(z)&=\frac{\epsilon_{ij}}{2}\int\frac{dk}{2\pi i}\tr\left[\Gamma \frac{z+H_{k\lambda}}{z^2-H_{k\lambda}^2}\partial_i H_{k\lambda}\,\frac{z+H_{k\lambda}}{z^2-H^2_{k\lambda}}\partial_jH_{k\lambda}\right]\\
&=\frac{\epsilon_{ij}}{2}\int\frac{dk}{2\pi i}\tr\left[\Gamma G_{k\lambda}(z)\partial_i H_{k\lambda}\,G_{k\lambda}(z)\partial_jH_{k\lambda}\right]\\
&=-{\epsilon_{ij}}\int\frac{dk}{4\pi i}\tr\left[\Gamma \partial_iG_{k\lambda}(z)\partial_jG^{-1}_{k\lambda}\right]\\
&=-{\epsilon_{ij}}\int\frac{dk}{4\pi i}\partial_i\tr\left[\Gamma G_{k\lambda}(z)\partial_jG^{-1}_{k\lambda}\right]=\partial_\lambda w_\lambda(z).
\end{align}
\end{subequations}
\end{widetext}
In the last line, we dropped total derivative of $k$ owing to the periodicity of the Hamiltonian in the present gauge.
Thus, $\partial_\lambda zF_\lambda(z)=\partial_\lambda w_\lambda(z)$ holds.

\subsection{Validity of the replacement in \Eq{eq:replace}}
\label{sec:decayA}
In this subsection, we show that the error of the replacement of the matrix element
\begin{equation}
\braket{R_1,\alpha|G_\lambda(z)\Gamma \partial_\lambda(G_\lambda(z)|R_2,\beta})
\end{equation}
in OBC with that in PBC is exponentially small.
Here, the lattice points $R_1$ and $R_2$ are assumed to be far from surfaces.
We first show this statement on the basis of the fact that $G_\lambda(z)$ $(\Im z\neq0)$ decays exponentially in space in PBC. We give an elementally proof of this fact in the next subsection.

For clarity, we write $G_\lambda(z)$ in PBC and OBC as $G_{\mathrm{PBC}}(z)$ and $G_{\mathrm{OBC}}(z)$, respectively.
Let us denote the difference of the Hamiltonian in PBC and OBC as $\hat{t}$, which includes matrix elements only between $R\sim0$ and $R\sim L$.
The left Dyson equation reads
\begin{equation}
G_{\mathrm{OBC}}(z)=G_{\mathrm{PBC}}(z)+G_{\mathrm{PBC}}(z)\,\hat{t}\,G_{\mathrm{OBC}}(z).
\end{equation}
Thus, we obtain
\begin{equation}
\bra{R_1,\alpha}\bigl\{G_{\mathrm{OBC}}(z)-G_{\mathrm{PBC}}(z)\bigr\}=\bra{R_1,\alpha}G_{\mathrm{PBC}}(z)\,\hat{t}\,G_{\mathrm{OBC}}(z),
\end{equation}
which is proportional to either $\braket{R_1,\alpha|G_{\mathrm{PBC}}(z)|R\sim0,\gamma}$ or $\braket{R_1,\alpha|G_{\mathrm{PBC}}(z)|R\sim L,\gamma}$. We conclude
\begin{equation}
\bra{R_1,\alpha}G_{\mathrm{OBC}}(z)=\bra{R_1,\alpha}G_{\mathrm{PBC}}(z)+O(e^{-\mathrm{min}[R_1,\,L-R_1]/\xi_0}),
\end{equation}
owing to the localization property of $G_{\mathrm{PBC}}(z)$ such that $\braket{R,\alpha|G_{\mathrm{PBC}}(z)|R',\beta}=O(e^{-|R-R'|/\xi_0})$ for $\Im z\neq0$ with some length scale $\xi_0$ independent of $L$.
The error term is negligible, for example, by assuming $R_c$ in Sec.~\ref{sec:derivation_of_SBBC}, and thus $R_1$, is $\sim L/2$.
We also obtain
\begin{equation}
G_{\mathrm{OBC}}(z)\ket{R_2,\beta}=G_{\mathrm{PBC}}(z)\ket{R_2,\beta}+O(e^{-\mathrm{min}[R_2,\,L-R_2]/\xi_0}),
\end{equation}
from the right Dyson equation. Thus, validity of the replacement has been established.

\subsection{Spatially decaying property of the Green's function}
Here, we give an elementally proof for spatially decaying property of Green's function for tight binding models.
In this section, we denote Green's function and Hamiltonian in PBC as $G(z)$ and $H$.
First, we can rewrite the matrix element of the Green's function as
\begin{equation}
\braket{R_1,\alpha|G(z)|R_2,\beta}=\int\frac{dk}{2\pi}\left(\frac{e^{ik\Delta R}}{z-H_k}\right)_{\alpha\beta}.
\end{equation}
Here, $H_k=\braket{k,\alpha|H|k,\beta}$ is a $2d\times2d$ matrix with $2d$ the number of internal degrees of freedom.
We can assume $\Delta R\equiv R_1-R_2>0$ without loss of generality, since the following discussion can be repeated with $k\to-k$.

Note that $H_{k+2\pi}=H_k$ in the present gauge. It means that we can take advantage of the Fourier expansion,
\begin{equation}
H_k=\sum_{l=-l_c}^{l_c}e^{ikl}h_{l}.
\end{equation}
The Fourier coefficient $h_l$ is a $2d\times 2d$ matrix, and the cutoff $l_c$ naturally follows from the reach of transfer integrals.
Let us define an analytic continuation
\begin{equation}
H(w)\equiv\sum_{l=-l_c}^{l_c}w^lh_{l},\label{eq:defHw}
\end{equation}
into $w\in\mathbb{C}$.
Then, we can write
\begin{equation}
\braket{R_1,\alpha|G(z)|R_2,\beta}=\oint_{|w|=1}\frac{dw}{2\pi i}\left(\frac{w^{\Delta R-1}}{z-H(w)}\right)_{\alpha\beta}.
\end{equation}
It is clear that the integrand is a meromorphic function of $w$, with \Eq{eq:defHw} and cofactor matrices of $z-H(w)$ in mind.
In particular, poles are absent on $|w|=1$, since we are considering the case $\Im z\neq0$ and hence $\det(z-H_k)\neq0$.
(This is also true for insulators when $z$ lies in the interval on the real axis corresponding to the band gap.)
Thus, we obtain for $\Delta R\to\infty$,
\begin{equation}
\braket{R_1,\alpha|G(z)|R_2,\beta}\to\left[\lim_{w\to w_0}\left(\frac{w-w_0}{z-H(w)}\right)_{\alpha\beta}\right]w_0^{\Delta R-1},\label{eq:realspaceGreen}
\end{equation}
where $w_0$ is the simple pole in the region $|w|<1$ with the largest absolute value.
It immediately follows that $\braket{R_1,\alpha|G(z)|R_2,\beta}\sim \exp(-|\Delta R|/\xi_0)$ with $\xi_0=-1/\ln|w_0|$. For example, $\xi_0$ is estimated to be $\sim 1/\ln|z|$  for $z$ with large $\Im z$.

We note that spatially decaying property of the density matrix of insulators readily follows from this result, by integrating \Eq{eq:realspaceGreen} on the contour encircling the interval on the real axis corresponding to occupied states.
It also means, along with the result of \Sec{sec:decayA}, matrix elements of the density matrix of insulators in OBC is asymptotically equivalent with that in PBC, as $R_1$ or $R_2$ get away from the surface. This fact is insightful for the asymptotic equivalence of the GWFs with the Wannier functions in PBC, when they are taken as the eigenstates of the projected position operator.
It follows that the real-space matrix elements of the projected position operators are almost equivalent in the bulk, and therefore so are their eigenstates\Cite{Vanderbilt2018,Kivelson1982}.

\subsection{Effect of surface disorders on SBBC}
\label{sec:impuritySBBC}
Here we comment on the effect of surface disorders on SBBC.
Note that the derivation of \Eq{eq:delF} remains valid even in the presence of surface disorders.
Therefore, the point is whether we can replace $H_\lambda$ and $G_\lambda(z)$ in OBC including surface disorders with those in the perfect crystal with PBC.
Here, it should be noticed that the discussion in Sec.~\ref{sec:decayA} holds even in the presence of surface disorders.
Actually, we can repeat the discussion with making $\hat{t}$ include surface disorders. Thus, contribution of surface disorders is exponentially small.

\end{document}